\documentclass[prd,twocolumn,showpacs]{revtex4}
\usepackage{graphicx}
\usepackage{latexsym}
\usepackage{amsmath}

\begin{document}

\title{Constraints on Dark Energy Models from Weak Gravity Conjecture\footnote{Supported by
the National Natural Science Foundation of China under Grant No 10605042}}
\author{CHEN Xi-Ming}
 \email{chenxm@cqupt.edu.cn}
 \affiliation{College of Mathematics and Physics, Chongqing University of Posts and Telecommunications,
Chongqing 400065, China}
\author{LIU Jie}
 \affiliation{College of Mathematics and Physics, Chongqing University of Posts and Telecommunications,
Chongqing 400065, China}
\author{GONG Yun-Gui}
 \email{gongyg@cqupt.edu.cn}
 \affiliation{College of Mathematics and Physics, Chongqing University of Posts and Telecommunications,
Chongqing 400065, China}
\begin{abstract}
We study the constraints on the dark energy model with constant
equation of state parameter $w=p/\rho$ and the holographic dark
energy model by using the weak gravity conjecture. The combination
of weak gravity conjecture and the observational data gives $w<-0.7$
at the $3\sigma$ confidence level. The holographic dark energy model
realized by a scalar field is in swampland.
\end{abstract}



\maketitle


The Type Ia supernova (SN Ia) observations discovered the
accelerated expansion of the Universe \cite{agr98} in 1998. The
observations of high redshift SN Ia with the Hubble Space Telescope
provide strong evidence for the transition from deceleration in the
past to acceleration at present \cite{riess06}. The current
accelerated expansion of the Universe is further supported by other
complementary astronomical observations, such as the Cosmic
Microwave Background (CMB) anisotropy measured by the Wilkinson
Microwave Anisotropy Probe (WMAP), the observations of the large
scale structure of the clusters of galaxies made by the Sloan
Digital Sky Survey (SDSS), etc \cite{astier,wmap3,sdss}. The energy
conditions were employed to derive the direct and model independent
evidence of the acceleration of the Universe \cite{gong07b,gong07c}.
To explain the late cosmic acceleration, an exotic dark component
with negative pressure, dubbed dark energy, is introduced. This in
turn imposes a big challenge to theoretical physics, because the
only observable effect of dark energy is through gravitational
interaction and its nature is still a total mystery. Although the
cosmological constant is the simplest candidate of dark energy which
is consistent with current observations, other possibilities are
also explored due to many orders of magnitude discrepancy between
the theoretical calculation and astronomical observations for the
vacuum energy. For a review of dark energy models, see \cite{DE}.

The direct way to explore the property of dark energy is through the
SN Ia, WMAP, SDSS and other observational data. One usually
parameterizes dark energy density or the equation of state parameter
$w(z)$ of dark energy to get some model independent information
about the property of dark energy
\cite{virey,sturner,gong06,gong07,astier01,huterer,weller,alam,gong04,gong05,lind,jbp,par1,par2,
par3,par4,par5,gong04a,wang05,jbp05,
nesseris6a,berger,sahni06,lihong,yun06,saini,jbp06,nesseris05}. On
the other hand, string theory is a promising theory of quantum
gravity. We expect string theory to give a resolution to the dark
energy problem. Recent progress in string theory suggests that
there are a large amount of landscape vacua in string theory. Vafa
argued that not all consistent looking effective field theories are
actually consistent \cite{vafa}. The large space of semi-classically
consistent effective field theories which are not consistent with
the full quantum theory is called swampland. Vafa applied the
finiteness of the number of massless fields to distinguish the
string landscape from swampland in \cite{vafa}. By observing that
gravity is the weakest force in nature, the authors in \cite{vafa1}
proposed a new criterion to distinguish parts of string landscape
from swampland. Basically, it says that for a four dimensional
$U(1)$ gauge theory with gauge coupling constant $g$, the intrinsic
UV cutoff is $\Lambda\sim g M_p$, where $M_p=(8\pi G)^{-1/2}$ is the
Planck mass. This is the weak gravity conjecture and it was further
studies in \cite{vafa2} with some evidences from string theory. Note
that the weak gravity here is referred to the weakness of gravity
compared with other forces in nature. In other words, it does not
mean that the strength of gravity is so weak that we can take linear
approximation. By applying this conjecture to scalar field, it was
suggested that the variation of the scalar field should be less than
the Planck energy scale, i.e., $\Delta\phi <M_p$ \cite{huang}.

For a scalar field $\phi$, the energy density is
$\rho=\dot{\phi}^2/2+V(\phi)$ and the pressure is
$\rho=\dot{\phi}^2/2-V(\phi)$. The equation of state parameter $w$
is
\begin{equation}
\label{sclrw}
w=\frac{\dot{\phi}^2/2-V(\phi)}{\dot{\phi}^2/2+V(\phi)}\ge -1.
\end{equation}
From the above equation, we find that $\dot{\phi}^2=(1+w)\rho$, so
the weak gravity conjecture tells us that
\begin{equation}
\label{weakgravconj}
\Delta\phi(z_m)=\int \dot\phi dt=\int_0^{z_m}\frac{\sqrt{(1+w)\rho}}{(1+z)H} dz<M_p,
\end{equation}
where $z_m$ is the highest redshift to apply the weak gravity
conjecture. It is clear that we may use this equation to constrain
the property of dark energy. The weak gravity conjecture was used to
discuss the equation of state parameter in \cite{huang1} and the
Chaplygin gas model in \cite{zhu}.

In this Letter, we apply the weak gravity conjecture
(\ref{weakgravconj}) to constrain the property of dark energy.
 In section II, we discuss the dark
energy model with constant $w$. We use the weak gravity
conjecture to constrain the matter energy density parameter
$\Omega_{m0}$ and $w$. Then we use the observational data to
constrain $\Omega_{m0}$ and $w$ and combine this result with the
constraint coming from the weak gravity conjecture. In section III,
we use the weak gravity conjecture to discuss the holographic dark
energy model.

\begin{figure}
\centering
\includegraphics[width=8cm]{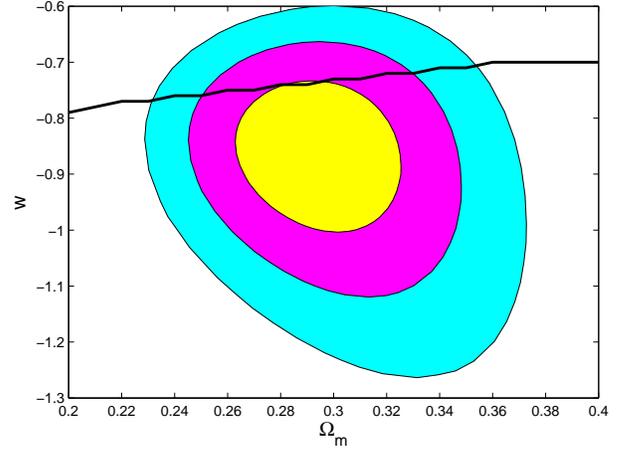}
\caption{The $1\sigma$, $2\sigma$ and $3\sigma$ contour constraints on $\Omega_m$
and $w$ from the combination of SN Ia data, the BAO measurement of SDSS and the shift parameter
from WMAP data. The area below the solid line is the allowed region from the weak gravity conjecture.}
\label{fig1}
\end{figure}

For the dark energy with constant equation of state parameter
$w=p/\rho$=constant, the potential of the scalar field is
\cite{wpotential}
\begin{eqnarray}
\label{sclrvphi}
V(\phi)&=&\left[\sqrt{\frac{\Omega_{m0}}{\Omega_{\phi 0}}}\sinh\left(-
\frac{3w}{2\sqrt{3(1+w)}}\frac{\phi-\phi_i}{M_p}\right)\right]^{2(1+w)/w}\nonumber \\
&&\times (1-w)\rho_{\phi 0}/2.
\end{eqnarray}
The Friedman equation is
\begin{equation}
\label{frweq}
H^2(z)=H^2_0\left[\Omega_{m0}(1+z)^3+(1-\Omega_{m0})(1+z)^{3(1+w)}\right],
\end{equation}
where the energy density parameter of matter $\Omega_m=\rho/3M_p^2
H^2$. The parameter of dark energy density is
\begin{equation}
\label{omegaw}
\Omega_\phi (z)=\frac{(1-\Omega_{m0})(1+z)^{3w}}{\Omega_{m0}+(1-\Omega_{m0})(1+z)^{3w}}.
\end{equation}
To apply the weak gravity conjecture, we need to evaluate the
expression
\begin{eqnarray}
\label{sclrweak1}
\frac{\Delta\phi(z_m)}{M_p}&=&\int_0^{z_m}\left(\frac{3(1+w)(1-\Omega_{m0})(1+z)^{3w}}{\Omega_{m0}
+(1-\Omega_{m0})(1+z)^{3w}}\right)^{1/2}\nonumber\\
&&\times (1+z)^{-1}dz,
\end{eqnarray}
here we take $z_m=1089$. To satisfy the weak gravity conjecture, we
require $\Delta\phi(1089)/M_p<1$, this gives constraints on the
parameters $\Omega_{m0}$ and $w$. The results are shown in figure
\ref{fig1}. In addition to this constraint, the observational data
can be used to constrain the parameters too. We use the 182 gold SN
Ia data given by Riess etal. \cite{riess06}, the baryon aoustic
oscillation (BAO) distance parameter \cite{sdss}
\begin{eqnarray}
\label{para1}
A&=&\frac{\Omega_{m}^{1/2}}{E^{1/3}(0.35)}\left(
\frac{1}{0.35} \int_0^{0.35}
\frac{dz}{E(z)}\right)^{2/3}\nonumber\\
&=&0.469(n/0.98)^{-0.35}\pm 0.17,
\end{eqnarray}
measured from the SDSS luminous red galaxy data , where $n=0.95$ as
determined from the WMAP three year data \cite{wmap3} and the
dimensionless Hubble pararmeter
\begin{eqnarray*}
E(z)&=&H(z)/H_0\\
&=&\left[\Omega_{m0}(1+z)^3+(1-\Omega_{m0})(1+z)^{3(1+w)}\right]^{1/2},
\end{eqnarray*}
and the shift parameter \cite{yun06}
\begin{equation}
\label{shift}
R=\Omega^{1/2}_{m0}\int_0^{1089} \frac{dz}{E(z)}=1.70\pm 0.03,
\end{equation}
as measured by the WMAP three year data \cite{wmap3}.

\begin{figure}
\centering
\includegraphics[width=8cm]{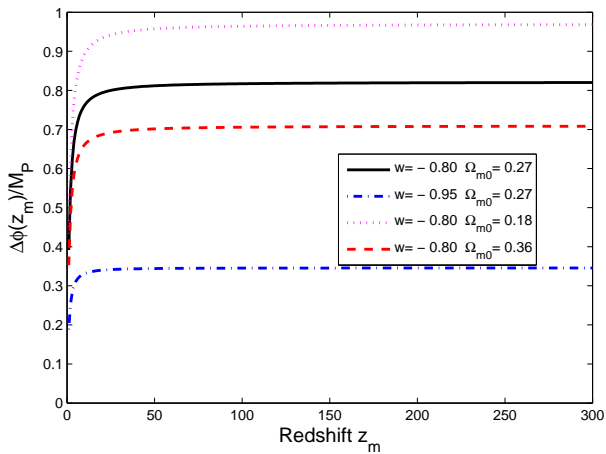}
\caption{$\Delta\phi(z_m)/M_p$ versus the maximum redshift $z_m$ for different $\Omega_{m0}$ and $w$.}
\label{fig2}
\end{figure}

For SN Ia data, we minimize
\begin{equation}
\label{chi}
\chi^2=\sum_i\frac{[\mu_{obs}(z_i)-\mu(z_i)]^2}{\sigma^2_i},
\end{equation}
where the extinction-corrected distance modulus
$\mu(z)=5\log_{10}[d_L(z)/{\rm Mpc}]+25$, $\sigma_i$ is the total
uncertainty in the SN Ia data, and the luminosity distance is
\begin{equation}
\label{lumdis}
d_L(z)=\frac{1+z}{H_0} \int_0^z
\frac{dz'}{E(z')},
\end{equation}
the nuisance parameter $H_0$ is marginalized over with flat prior.
For the BAO distance parameter, we add the term
$$\left[\frac{A-0.469(0.95/0.98)^{-0.35}}{0.017}\right]^2$$
to $\chi^2$. For the shift parameter, we add the term
$$\left(\frac{R-1.70}{0.03}\right)^2$$
to $\chi^2$. The $1\sigma$, $2\sigma$ and $3\sigma$ contour
constraints on $\Omega_{m0}$ and $w$ are shown in figure \ref{fig1}.
From figure \ref{fig1}, we see that the weak gravity conjecture can
further reduce the parameter space. At $3\sigma$ level, $w<-0.7$.
As we mentioned above, we choose to integrate $d\phi$ to $z_m=1089$.
In fact the choice of this redshift is high enough, we can choose
smaller $z_m$ because $\Omega_\phi \ll 1$ when $z\gg 1$ for most
dark energy models, so the contribution of higher redshift is
negligible. To confirm this, we plot $\Delta\phi/M_p$ versus $z_m$
in figure \ref{fig2} for different $\Omega_m$ and $w$. From figure
\ref{fig2}, we see that $\Delta\phi(z_m)/M_p$ is saturate when $z_m$
reaches $z=40$. We also see that the value of $\Delta\phi/M_p$
becomes smaller as we decrease $w$ or increase $\Omega_{m0}$. This
is easily understood because $\Omega_\phi$ becomes smaller.

\begin{figure}
\centering
\includegraphics[width=8cm]{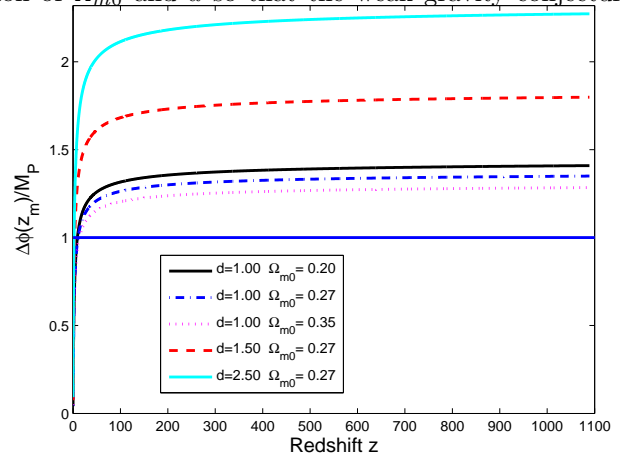}
\caption{$\Delta\phi(z_m)/M_p$ versus the maximum redshift $z_m$ for different $\Omega_{m0}$ and $d$.}
\label{fig3}
\end{figure}

For the holographic dark energy model, the energy density is
\cite{li}
\begin{equation}
\label{rhod}
\rho_\Lambda = 3 d^2 M^2_p R^{-2}_h,
\end{equation}
where the proper size of the future event horizon $R_h$ is
\begin{equation}
\label{rheq}
R_h(t)=a(t)\int^\infty_t \frac{dt'}{a(t')}.
\end{equation}
The evolution equation of $\Omega_\Lambda=\rho_\Lambda/(3M^2_p H^2)$
is
\begin{equation}
\label{omegadeq}
\frac{d\Omega_\Lambda}{dz}=-\frac{1}{1+z}\Omega_\Lambda(1-\Omega_\Lambda)
\left(1+\frac{2}{d}\sqrt{\Omega_\Lambda}\right).
\end{equation}
The solution of the above equation (\ref{omegadeq}) is
\begin{eqnarray}
\label{omegadzeq}
\ln \Omega_\Lambda - {d \over 2+d} \ln (1 - \sqrt{\Omega_\Lambda})
+ {d \over 2-d} \ln (1 +\sqrt{\Omega_\Lambda})\nonumber\\
 - {8 \over 4-d^2} \ln(d +
2 \sqrt{\Omega_\Lambda}) = - \ln(1 + z) + y_0,
\end{eqnarray}
 where $y_0$ can be
determined by the value of $\Omega_{\Lambda 0}$ through equation
(\ref{omegadzeq}). The equation of state parameter is
\begin{equation}
\label{wdeq}
w_\Lambda=-\frac{1}{3}\left(1+\frac{2}{d}\sqrt{\Omega_\Lambda}\right).
\end{equation}
To get a scalar field model corresponding the above model, we must
require $d\ge 1$ so that $w_\Lambda\ge -1$. To apply the weak
gravity conjecture, we need to calculate
\begin{eqnarray}
\label{sclrweak2}
\frac{\Delta\phi(z_m)}{M_p}&=\int_0^{z_m}[3(1+w_\Lambda)\Omega_\Lambda]^{1/2}d\ln(1+z)\nonumber\\
&=\int_0^{z_m}\left[2\left(\Omega_\Lambda-\frac{\Omega_\Lambda^{3/2}}{d}\right)\right]^{1/2}d\ln(1+z).
\end{eqnarray}
Substituting the solution (\ref{omegadzeq}) into the above integration
(\ref{sclrweak2}), we obtain the result $\Delta\phi(z_m)/M_p$. Again,
we take $z_m=1089$ which is enough as reasons discussed above.
We plot the results of $\Delta\phi(z_m)/M_p$
versus the maximum redshift $z_m$ in figure \ref{fig3}. It is clear
from figure \ref{fig3} that the value of $\Delta\phi(z_m)/M_p$
saturates before the redshift $z=1089$. In this case, we cannot find
any combination of $\Omega_{m0}$ and $d$ so that the weak gravity
conjecture is satisfied. In other words, the holographic dark energy
model realized by a scalar field is in the swampland.
The larger the value of $\Omega_{m0}$ or $d$, the smaller the value of
$\Delta\phi(z_m)/M_p$. When we increase $d$, $w$  also increases, so
$\Omega_\Lambda$ decreases faster.

In summary, we have applied the weak gravity conjecture to discuss the
dark energy model with constant $w$ and the holographic dark energy
model. For the dark energy model with constant $w$, we also used the
SN Ia data, the BAO distance parameter and the shift parameter to
constrain the parameters $\Omega_{m0}$ and $w$, and we find that
$w<-0.7$ at the $3\sigma$ confidence level. In \cite{huang1}, the
author also discussed the dark energy model with constant $w$ by
using the weak gravity conjecture. However, the author fixed
$\Omega_{m0}$ and did not consider the degeneracies between
$\Omega_{m0}$ and $w$. Our work not only consider the more general
case, but also used the observational data to constrain the
parameter space of $\Omega_{m0}$ and $w$. For the holographic dark
energy model, we find that it is in swampland. We also discussed the
choice of the highest redshift $z_m$ for the integration and find
that the integration in (\ref{weakgravconj}) saturates before
$z_m=1089$, so it is enough to choose $z_m=1089$. Furthermore, for
some cases, we can choose much smaller redshift.

\end{document}